\documentclass{ifacconf}
\usepackage{enumerate}
\usepackage{amssymb}
\usepackage{amsmath,amssymb,amsfonts,mathrsfs}
\usepackage{amsmath}
\usepackage[latin1]{inputenc}
\usepackage{mathrsfs}
\usepackage{psfrag}
\usepackage{epsfig}
\usepackage{graphicx}
\usepackage{amsfonts} 
\usepackage[usenames,dvipsnames,svgnames,table]{xcolor}
\usepackage{array}
\usepackage{tikz}
\usepackage{booktabs}
\usepackage{multirow}
\usepackage{natbib} 
\usepackage{flushend}

\def\qedp{\hspace*{\fill}~{\tiny $\blacksquare$}}

\def\qed{\relax\ifmmode\hskip2em \Box\else\unskip\nobreak\hskip1em $\Box$\fi}

\newtheorem{theorem}{Theorem}

\newtheorem{itdefinition}{Def.}
\newtheorem{itproposition}{Proposition}
\newtheorem{itresult}{Result}
\newtheorem{itremark}{Remark}
\newtheorem{itassumption}{Assumption}
\newtheorem{itcorollary}{Corollary}
\newtheorem{itexample}{Example}

\newenvironment{assumption}{\begin{itassumption}\rm}{\end{itassumption}}

\begin{document}

\begin{frontmatter}

\title{
Designing Experiments for Data-Driven Control of Nonlinear Systems
} 


\author[First]{Claudio De Persis} 
\author[Second]{Pietro Tesi} 

\address[First]{ENTEG, University of Groningen,  Nijenborgh 4, 9747AG 
   Groningen, The Netherlands (e-mail: c.de.persis@rug.nl).}
\address[Second]{DINFO, University of Florence, 50139 Florence, Italy (e-mail: pietro.tesi@unifi.it)}

\begin{abstract}                
In a recent paper we have shown that data collected from linear systems excited by persistently exciting inputs during \textcolor{black}{low-complexity} experiments, can be used to design state- and output-feedback controllers, including optimal Linear Quadratic Regulators (LQR),  by solving linear matrix inequalities (LMI) and  semidefinite programs. We have also shown how to stabilize in the first approximation unknown nonlinear systems using data.  In contrast to the case of linear systems, however, in the case of nonlinear systems the conditions for \textcolor{black}{learning a controller directly from data} may not be fulfilled even when the data are collected in experiments performed using persistently exciting inputs. In this paper we show how to design experiments that lead to the fulfilment of these conditions. 
\end{abstract}


\end{frontmatter}

\section{Introduction}

Recent advances in learning have prompted a renewed interest in the use of data for control of complex dynamical systems in at least two different ways. On one hand, the use of new learning-based identification techniques for system identification followed by  off-the-shelf top-notch robust control design \citep{chiuso2016regularization,recht2018tour}. On the other hand, the design of control policies directly form data, skipping altogether any  attempt of identifying the system's model \citep{campi2002virtual,campi2006direct,formentin2013optimal,novara16data-driven,tanaskovic2017data}. Related contributions are the iterative feedback tuning \citep{Hjalmarsson1998}, correlation-based tuning \citep{karimi2007non}, and the design of controllers for all the systems compatible with the measured data via polynomial optimization  
\citep{Dai2018}. Optimal \citep{Goncalves2019, baggioCSL19} and nonlinear \citep{wabersich2018ecc} control problems have also been investigated. 

Inspired by the the so-called {\em Fundamental Lemma} of \cite{willems2005note},  a novel approach to simulate and control unknown dynamical systems without any identification, but merely using finite-length input-output data, was introduced in  \cite{Markovsky2008}. The spotlight on  the ideas of  \cite{markovsky2005algorithms,Markovsky2008} and their important role in data-driven control design was turned on again by the recent paper by \cite{Coulson19ecc}. See \cite{Coulson19cdc} and \cite{Huang19cdc} for recent follow-ups.

Also relying on the results of  \cite{willems2005note}, recently \cite{cdptCDC,depersis-tesi2020tac} have shown that data collected from linear systems excited by persistently exciting inputs during low-complexity finite-horizon off-line experiments, can be used to design state- and output-feedback controllers, including optimal Linear Quadratic Regulators (LQR),  by solving linear matrix inequalities (LMI) and  semidefinite programs. The authors have also shown how these methods are robust to the use of  data corrupted by bounded-but-unknown deterministic noise,  meaning that they return a stabilizing controller for noisy data that satisfy a quantified signal-to-noise ratio.   

{\color{black}In a continuation of \cite{cdptCDC,depersis-tesi2020tac},} the necessity of the persistency-of-excitation conditions has been throughly discussed in \cite{vanwaarde2019data}, the design of controllers robust to process disturbances has been further explored in \cite{berberich2019robust} and data-based guarantees for polyhedral set-invariance properties (safe controllers) have been studied in \cite{bisoffi19ifac}. 

Another result of \cite{depersis-tesi2020tac} was to extend the stabilization result to the case of nonlinear systems (see Theorem \ref{prop:nonlinear_stability} in Section \ref{sec:prelim} below). Namely, if the system is nonlinear and if the data collected during a finite-length off-line experiment satisfy suitable conditions (Assumptions \ref{ass:dist} and \ref{ass:dist2} below) then we can design a controller from these data that stabilizes the system in the first approximation. In contrast with what happens with linear systems, however, where it can be shown that these conditions are always satisfied if  persistently exciting inputs  are applied in the experiment, in the case of nonlinear systems the fulfilment of Assumptions \ref{ass:dist} and \ref{ass:dist2} is a more challenging task.

The purpose of  this note is threefold. First we show that  experiments that are performed on nonlinear systems  may fail to generate data that satisfy  Assumptions \ref{ass:dist} and \ref{ass:dist2}, even when persistently exciting inputs are applied.  Second, we show how to design experiments that lead to the fulfilment of these assumptions. Third, we provide some heuristic considerations on how to verify that the stabilizing controller that we design is actually stabilizing when no a priori knowledge about the nonlinear system is available. 

In Section \ref{sec:prelim} we recall the results of \cite{depersis-tesi2020tac}. The main results are in Section \ref{sec:main}. Conclusions are drawn in Section \ref{sec:conclusion}.

\section{Preliminaries}\label{sec:prelim}

We start by recalling the result given in \cite{depersis-tesi2020tac}.

Consider a smooth nonlinear system 
\begin{equation}
\label{nonl}
x(k+1) = f(x(k),u(k))
\end{equation}
and let $(\overline x, \overline u)$ be a \emph{known} equilibrium pair, 
that is such that 
$\overline x = f(\overline x, \overline u)$.
Throughout his note, for ease of notation we will assume without loss
of generality that the equilibrium point of interest is $(\overline x,\overline u)=(0,0)$. 

Let us rewrite the nonlinear system as
\begin{eqnarray} \label{nonl2}
 x(k+1) = A  x(k)  + B  u(k) + d(k)
\end{eqnarray}
where
\begin{eqnarray} \label{lin_matrices}
A:=\left.\frac{\partial f}{\partial x}\right|_{(x,u)=(0, 0)},\quad
B:=\left.\frac{\partial f}{\partial u}\right|_{(x,u)=(0, 0)} \,.
\end{eqnarray}
The quantity $d$ accounts for higher-order terms and it has the property that it 
goes to zero faster than $x$ and $u$, namely we have
\begin{eqnarray*} 
d=R( x,  u)\begin{bmatrix}  x \\  u\end{bmatrix}
\end{eqnarray*}
with $R( x,  u)$ a  matrix of smooth functions with the property that 
\begin{eqnarray} \label{eq:convergence_d}
\lim_{{\tiny \begin{bmatrix}  x \\  u\end{bmatrix}\to {0}}} R( x,  u)= {0}
\end{eqnarray}

It is known that if the pair $(A,B)$ defining the linearized 
system is stabilizable then the controller $K$ rendering $A+BK$ 
stable exponentially stabilizes the equilibrium of the nonlinear system. 
Let  
\begin{eqnarray*}
X_{0, T} &:=& \left[ \begin{array}{cccc}  x(0) &  x(1) & \cdots &  x(T-1) \end{array} \right] \\
X_{1, T} &:=& \left[ \begin{array}{cccc}  x(1) &  x(2) & \cdots &  x(T) \end{array} \right]  \\
U_{0, T} &:=& \left[ \begin{array}{cccc}  u(0) &  u(1) & \cdots &  u(T-1) \end{array} \right] \\
D_{0, T} &:=& \left[ \begin{array}{cccc} d(0) & d(1) & \cdots & d(T-1) \end{array} \right]
\end{eqnarray*} 
be data resulting from an experiment carried out on the nonlinear system \eqref{nonl}.
Note that the matrices $X_{0, T}$, $X_{1, T}$ and 
$U_{0, T}$ are available from data. 

Consider the following assumptions.

\smallskip 
\begin{assumption} \label{ass:dist}
The matrices 
 \begin{eqnarray}
\begin{bmatrix}
U_{0,T} \\
\hline
X_{0, T}
\end{bmatrix}, \;X_{1, T}
\end{eqnarray} 
have full row rank. 
\smallskip

\end{assumption}
\begin{assumption} \label{ass:dist2}
It holds that 
 \begin{eqnarray}
D_{0, T} D_{0, T}^\top \preceq \gamma X_{1, T} X_{1, T}^\top
\end{eqnarray} 
for some $\gamma >0$. 
\end{assumption} 
\smallskip

The following result holds.  

\smallskip
\begin{theorem}[\cite{depersis-tesi2020tac}] \label{prop:nonlinear_stability} 
Consider a nonlinear system as in \eqref{nonl}, 
along with an equilibrium pair $(\overline x, \overline u)$.
Suppose that  
Assumptions \ref{ass:dist} and \ref{ass:dist2} hold.
Then, any solution $(Q,\alpha)$ to
\begin{eqnarray} \label{lmi.state.feedback.nonlinear}
\begin{array}{l}
\max_{(Q,\alpha)} \alpha \\[0.4cm] 
\textrm{subject to} \\[0.3cm]
\begin{array}{rl}
\begin{bmatrix}
X_{0,T} \,Q - \alpha X_{1,T}  X_{1,T}^\top &  X_{1,T} Q\\
Q^\top X_{1,T}^\top & X_{0,T} \,Q
\end{bmatrix}
\succeq & 0\\[4mm]
\begin{bmatrix}
I_T & Q \\
Q^\top & X_{0,T} \,Q
\end{bmatrix}
\succeq & 0 
\end{array}
\end{array}
\end{eqnarray}
such that $\gamma < \alpha^2/(4 + 2\alpha)$, $\alpha>0$, returns a 
state-feedback controller
$K = U_{0, T}Q (X_{0,T} Q )^{-1}$ that locally stabilizes 
the equilibrium $(\overline x, \overline u)$. \qedp
\end{theorem}

The reason for considering the maximization of $\alpha$ is to render condition 
$\gamma < \alpha^2/(4 + 2\alpha)$ easier to fulfil. Notice that (\ref{lmi.state.feedback.nonlinear})
is a semidefinite program.

Theorem \ref{prop:nonlinear_stability} rests on the fact that
Assumptions \ref{ass:dist} and \ref{ass:dist2} hold and that (\ref{lmi.state.feedback.nonlinear})
admits a solution with $\gamma < \alpha^2/(4 + 2\alpha)$. 
For \emph{linear} controllable systems, this is always the case provided that the 
experiments are carried out with persistently
exciting inputs \textcolor{black}{of a sufficiently high order (see
\cite[Definition 1]{depersis-tesi2020tac} for a definition of persistency of excitation).}
In fact, under these conditions: (i) Assumptions \ref{ass:dist} and \ref{ass:dist2} are satisfied;
(ii) problem (\ref{lmi.state.feedback.nonlinear}) 
is feasible, and any solution is such that $\gamma < \alpha^2/(4 + 2\alpha)$
(since $D_{0, T}=0$  
Assumption \ref{ass:dist2} holds with an arbitrary $\gamma$).
We summarize this fact. 

\begin{theorem}[\textcolor{black}{\cite{depersis-tesi2020tac}}] \label{thm:linear} 
Let (\ref{nonl}) be a {\em linear} controllable system 
with \textcolor{black}{state-space dimension $n$}, and consider an experiment carried out
with a persistently exciting input \textcolor{black}{of order $n+1$}. Then, 
Assumptions \ref{ass:dist} and \ref{ass:dist2} hold.
Further, problem (\ref{lmi.state.feedback.nonlinear}) is feasible
and any solution is such that $\gamma < \alpha^2/(4 + 2\alpha)$.
\textcolor{black}{Thus, any solution returns a
stabilizing state-feedback gain
$K = U_{0, T}Q (X_{0,T} Q )^{-1}$. \qedp
}
\end{theorem}

In the sequel, it will be useful to relate the 
evolution of the nonlinear system with the one of the corresponding linearized system, 
which we express as
\begin{eqnarray} \label{lin}
x^l(k+1) =  A x^l(k) + B u^l(k) 
\end{eqnarray}
We also let
\begin{eqnarray*}
X^l_{0, T} &:=& \left[ \begin{array}{cccc}  x^l(0) &  x^l(1) & \cdots &  x^l(T-1) \end{array} \right] \\
X^l_{1, T} &:=& \left[ \begin{array}{cccc}  x^l(1) &  x^l(2) & \cdots &  x^l(T) \end{array} \right]  \\
U^l_{0, T} &:=& \left[ \begin{array}{cccc}  u^l(0) &  u^l(1) & \cdots &  u^l(T-1) \end{array} \right] 
\end{eqnarray*} 
be the data resulting from an (hypothetical) experiment made on the linearized system \eqref{lin}.
Note that the matrices $X^l_{0, T}$, $X^l_{1, T}$ and 
$U^l_{0, T}$ are not available from data.

\section{Main results}\label{sec:main}

Since around an equilibrium point nonlinear systems behave as linear systems, 
it is tempting to conclude that Theorem \ref{thm:linear} can be extended to the nonlinear case 
as long as the experiments are carried out sufficiently close
to the equilibrium point. As we will see, 
this is true \emph{only if} the experiments are carried out in  a certain manner.
More precisely, we will see that there indeed exist experiments that ensure the same
properties as in Theorem \ref{thm:linear},
but not all experiments (even with persistently exciting inputs
and arbitrarily close to the
equilibrium) guarantee these properties.
In the remainder of this section:
\begin{enumerate}
\item we show that there exist experiments (carried out with persistently exciting inputs
and arbitrarily close to the
equilibrium) for which Assumptions \ref{ass:dist} and \ref{ass:dist2} fail to hold;
\item we show that there exist experiments (carried out with persistently exciting inputs
and sufficiently close to the
equilibrium) ensuring the same
properties as in Theorem \ref{thm:linear}, and we characterize 
a class of experiments ensuring these properties;
\item we provide some heuristic considerations on how to verify that a
solution to (\ref{lmi.state.feedback.nonlinear}) returns
a stabilizing controller. (In general, one cannot determine how close to the equilibrium 
the experiments should be run unless we have some prior knowledge on $D_{0, T}$,
that is on the type of nonlinearity; thus, in general one cannot assess the fulfilment of the
condition $\gamma < \alpha^2/(4 + 2\alpha)$ since $\gamma$ is unknown.)
\end{enumerate}

\subsection{Experiment design issues for nonlinear systems}

We start with showing that there exist experiments (carried out with persistently exciting inputs
and arbitrarily close to the
equilibrium) for which Assumptions \ref{ass:dist} and \ref{ass:dist2} fail to hold.
We show this fact through an example.

\textcolor{black}{\textbf{Example}}.
Consider the nonlinear system
\[
x(k+1) = x(k)^2 +u(k)
\]

with equilibrium $(\overline x,\overline u)=(0,0)$. 
The system can be 
written as
\[
 x(k+1) =  u(k) + d(k), \quad d(k) = x(k)^2
\]

with $A=0$ and $B=1$, 
and the corresponding linearized system is given by
\[
x^l(k+1) =  u^l(k)
\]

Consider now the experiment given by
\begin{eqnarray*}
&& x(0) = x_0 \\
&& U_{0, T} = 
\left[
\begin{array}{ccc} u(0) & u(1) & u(2) \end{array}
\right]
\end{eqnarray*}

where 
\[
x_0 = u(0) = \theta, \,\, u(1) = \theta + \theta^2, \,\, 
u(2) = \theta + \theta^2 + (\theta + \theta^2)^2
\]
with \textcolor{black}{$\theta$ real.}
The input is persistently exciting of order $n+1=2$. 
Now, If we hypothetically apply this experiment to the linearized system,
that is we let $x^l(0)=x_0$ and $\textcolor{black}{U^l_{0, T} = U_{0, T}}$,
we obtain 
\[
\left[
\begin{array}{c} U^l_{0,T} \\ X^l_{0,T}  \end{array}
\right]  =
\left[
\begin{array}{ccc} \theta & \theta + \theta^2 & \theta + \theta^2 + (\theta + \theta^2)^2 \\
\theta & \theta & \theta + \theta^2 \end{array}
\right]
\]
which is full row rank
for all $\theta \neq 0$, and the same holds for $X^l_{1,T}$. However,
\[
\begin{array}{rl}
\left[
\begin{array}{c} U_{0,T} \\ X_{0,T}  \end{array}
\right]  =&
\left[
\begin{array}{ccc} u(0) & u(1) & u(2) \\
x(0) & u(0) + x(0)^2 & u(1) + x(1)^2
\end{array}
\right] \\[4mm]
=&
\left[
\begin{array}{ccc} \theta & \theta + \theta^2 & \theta + \theta^2 + (\theta + \theta^2)^2 \\
\theta & \theta + \theta^2 & \theta + \theta^2 + (\theta + \theta^2)^2 \end{array}
\right]
\end{array}
\]
has rank $1$ for all $\theta \neq 0$. We conclude that
Assumption \ref{ass:dist} does not hold. Note that this holds true
even though the input sequence if persistently exciting and $\theta$ 
is arbitrarily small. \qedp

This simple example shows that for nonlinear systems 
there might exist experiments for which Assumption \ref{ass:dist} is not satisfied, even if
these experiments originate from persistently exciting inputs and are carried out 
arbitrarily close to the equilibrium.

An intuitive explanation for this fact can be obtained by relating the matrices 
associated with the linear and nonlinear system. 
Consider an experiment on the nonlinear system with initial condition $x_0$ and input sequence 
\textcolor{black}{$\{u(0),u(1),\ldots,u(T-1)\}$} resulting in state matrices $X_{0,T}$ and $X_{1,T}$.
It is simple to see that 
\[
x(k) = x^l(k) + \sum_{i=0}^{k-1} A^{k-i-1} d(i)
\]
This implies that the matrices in Assumption \ref{ass:dist} can be written as 
 \begin{eqnarray} \label{eq:main0}
\begin{bmatrix}
U_{0,T} \\
\hline
X_{0, T}
\end{bmatrix} =
\begin{bmatrix}
U^l_{0,T} \\
\hline
X^l_{0, T}
\end{bmatrix} + 
\begin{bmatrix}
0 \\
\hline
\Xi
\end{bmatrix}, \,\,
X_{1, T} = X^l_{1, T} + \Psi
\end{eqnarray} 
where 
\begin{equation} \label{eq:Xi}
\Xi := 
\begin{bmatrix}
0 & d(0) & A d(0) + d(1) & A^2 d(0) + A d(1) + d(2) & \cdots & 
\end{bmatrix}
\end{equation}
and
\begin{equation} \label{eq:Psi}
\Psi := 
\begin{bmatrix}
d(0) & A d(0) + d(1) & A^2 d(0) + A d(1) + d(2) & \cdots & 
\end{bmatrix}
\end{equation} 
In connection with the previous example, this translates to
\[
\begin{array}{rl}
\left[
\begin{array}{c} U^l_{0,T} \\ X^l_{0,T}  \end{array}
\right]  = &
\left[
\begin{array}{ccc} \theta & \theta + \theta^2 & \theta + \theta^2 + (\theta + \theta^2)^2 \\
\theta & \theta & \theta + \theta^2 \end{array}
\right]\\[4mm]
\begin{bmatrix}
0 \\ \Xi 
\end{bmatrix}
= &
\begin{bmatrix}
0 & 0 & 0 \\
0 & \theta^2 & (\theta + \theta^2)^2 
\end{bmatrix}
\end{array}
\]

This means that 
\textcolor{black}{there exist trajectories (i.e. experiments) for which
the perturbation that causes 
\begin{equation} \label{eq:lin_mat} 
\left[
\begin{array}{c} U^l_{0,T} \\ X^l_{0,T}  \end{array}
\right]
\end{equation} 
to lose rank is of the same order ($\theta^2$ in this example) as the nonlinear terms, thus  
of the same order as $\Xi$. In the sequel, we will 
show that one can nonetheless design experiments 
so as to satisfy Assumptions \ref{ass:dist} and \ref{ass:dist2}. 
}

\subsection{Scaling the experiments ensures Assumptions \ref{ass:dist} and \ref{ass:dist2}}

The basic idea is to show that there also exist experiments such that $\Xi$
vanishes faster than the 
perturbation causing the matrix (\ref{eq:lin_mat}) to lose rank. 

Consider an experiment with initial condition $x_0$ and input sequence 
 \[
u_{0,T} = \{u(0),u(1),\ldots,u(T-1)\}
\]
persistently exciting of order $n+1$. For this experiment, relation (\ref{eq:main0}) holds.
Consider now a \emph{scaled} version of this experiment with initial condition
$\varepsilon x_0$ and input sequence $\varepsilon u_{0,T}$
with $\varepsilon > 0$ real. Denote by $\tilde U_{0,T}$ the  corresponding 
matrix which satisfies
\begin{eqnarray*}
\tilde U_{0,T} &=& 
\begin{bmatrix}
\tilde u(0) & \tilde u(1) & \cdots & \tilde u(T-1)
\end{bmatrix} \\
&=& \varepsilon U_{0,T} = \varepsilon U^l_{0,T}
\end{eqnarray*}
Also, denote by $\tilde X_{0, T}$ and $\tilde X_{1, T}$ the state matrices of the nonlinear system 
resulting from this new experiment, 
\[\begin{array}{rl}
\tilde X_{0,T} = &
\begin{bmatrix}
\tilde x(0) & \tilde x(1) & \cdots & \tilde x(T-1)
\end{bmatrix}\\[0.5cm]
\tilde X_{1,T} = &
\begin{bmatrix}
\tilde x(1) & \tilde x(2) & & \cdots & \tilde x(T)
\end{bmatrix}
\end{array}
\]
where $\tilde x$ satisfies
\begin{eqnarray*}
\tilde x(k+1) &=& f (\tilde x(k),\tilde u(k)) \\ &=& A  \tilde x(k)  + B \tilde x(k)  + \tilde d(k) 
\end{eqnarray*}
with $\tilde x(0) = \varepsilon x_0$,
and where, as before,
\begin{eqnarray*} 
\tilde d= R(\tilde x, \tilde u)
\begin{bmatrix} \tilde x \\ \tilde u \end{bmatrix}
\end{eqnarray*}
with $R(\tilde x, \tilde u)$ a matrix of smooth functions with the property that 
\begin{eqnarray} \label{eq:convergence_d}
\lim_{{\tiny \begin{bmatrix} \tilde x \\ \tilde u\end{bmatrix}\to {0}}} R(\tilde x, \tilde u)= {0}
\end{eqnarray} 

Now, for this new experiment it holds that
\begin{eqnarray} \label{eq:main1}
\begin{bmatrix}
\tilde U_{0,T} \\
\hline
\tilde X_{0, T}
\end{bmatrix} = \varepsilon
\begin{bmatrix}
 U^l_{0,T} \\
\hline
 X^l_{0, T}
\end{bmatrix} + 
\begin{bmatrix}
0\\
\hline
\tilde \Xi
\end{bmatrix}, \,\,
\tilde X_{1, T} = \varepsilon X^l_{1, T} + \tilde \Psi
\end{eqnarray} 
where the matrices $\tilde \Xi$ and $\tilde \Psi$ are defined as in (\ref{eq:Xi}) and 
(\ref{eq:Psi}) with $d$ replaced by $\tilde d$. 
We will only show the first of (\ref{eq:main1}) since the reasoning for the second is analogous.

In order to prove the first of (\ref{eq:main1}) it is sufficient to show that 
$\tilde X_{0, T} = \varepsilon X^l_{0, T} + \tilde \Xi$ since the 
relation $\tilde U_{0,T} = \varepsilon U^l_{0,T} $ holds by construction. 
The result can be proved by induction on the entries $\tilde x$ of the matrix
$\tilde X_{0, T}$. The claim holds for $k=0$.
Suppose that the claim holds up to a certain $k \geq 0$. We have
\begin{eqnarray}
\tilde x(k+1) &=& A \tilde x(k) + B \tilde u(k) + \tilde d(k) \nonumber \\
&=& A \left( \varepsilon x^l(k) + \sum_{i=0}^{k-1} A^{k-i-1} \tilde d(i)  \right) + B \tilde u(k) + \tilde d(k) \nonumber \\
&=& \varepsilon \left( A x^l(k) + B  u^l(k) \right) + \sum_{i=0}^{k-1} A^{k-i} \tilde d(i) + \tilde d(k) \nonumber \\
&=& \varepsilon x^l(k+1) + \sum_{i=0}^{k} A^{k-i} \tilde d(i)  
\end{eqnarray}
which gives the claim.

The important fact resulting from these relations is that the matrices $X^l_{0, T}$ and 
$X^l_{1, T}$ are \emph{fixed}, and the perturbation which cause them to lose rank
depends on $\varepsilon$. To this end, we recall the following result.
\smallskip

\begin{theorem} \label{eq:rank_pert}
{\rm \cite[Theorem 5.1]{dahleh}}
Suppose $M \in \mathbb C^{m \times n}$ is full column rank. Then,
\[
\min_{\Delta \in \mathbb C^{m \times n}}
\left\{  \| \Delta \| \,\, | \,\, M+\Delta \textrm{ has rank} < n
\right\} = \sigma_n(M)
\]
where $\| \cdot \|$ denotes the Euclidean norm, and where $\sigma_n$ is the smallest singular 
value of the matrix $M$. \qedp
\end{theorem}

The reasoning 
for full row rank matrices is analogous (one can alternately consider (\ref{eq:main1}) with the transpose).

Recalling that the singular values of a matrix $M$ are the square roots of the eigenvalues of the matrix
$M M^\top$, in connection with (\ref{eq:main1}) it follows that the smallest perturbation 
that causes $\varepsilon M$ with
\[
M := 
\begin{bmatrix}
 U^l_{0,1,T} \\
\hline
 X^l_{0, T}
\end{bmatrix}
\]
to lose rank is of order $\varepsilon$. In fact, since $M$ is full row rank then the smallest
eigenvalue of the matrix $M M^\top$ is strictly positive. Denoting this eigenvalue by $\underline \lambda$,
it follows that the smallest singular value of $\varepsilon M$ is $\varepsilon \sqrt{\underline \lambda}$.
Hence, 
to prove that the perturbed matrix in \eqref{eq:main1} is full row rank,
it is sufficient to show that the matrix $\tilde \Xi$
goes to zero faster than $\varepsilon$ since this implies the existence of a value $\overline \varepsilon$ 
such that
\begin{equation}\label{perturbation.bound}
\| \tilde \Xi \| < \varepsilon \sqrt{\underline \lambda} \quad \forall \varepsilon \in (0, \overline \varepsilon)
\end{equation}

This fact can be proven by looking at the various elements $\tilde d$ which form  
the entries of the matrix $\tilde \Xi$, 
since 
\small
\[
\| \tilde \Xi \|^2 \le \displaystyle \sum_{k=1}^{T-1} {\rm trace} 
\left[
\left(\sum_{i=0}^{k-1} A^{k-1-i} \tilde d(i)\right)^\top
\left(\sum_{i=0}^{k-1} A^{k-1-i} \tilde d(i)\right)
\right]
\]
\normalsize

with the right-hand side being the Frobenius norm $\| \tilde \Xi \|_F^2$.  
In fact, if one can prove that $\tilde d$ converges to zero faster than $\varepsilon$, then the bound \eqref{perturbation.bound} on the maximum allowable perturbation holds. 
\smallskip

\textbf{Claim}. For all $k\ge 0$, $\tilde d(k)$ converges to zero faster than $\varepsilon$.

\smallskip

{\it Proof of the Claim.} Recall that, for all $k$, 
\begin{equation}\label{d.tilde} 
\tilde d(k) = R(\tilde x(k), \tilde u(k))
\begin{bmatrix} \tilde x(k) \\ \tilde u(k) \end{bmatrix}
\end{equation}
with $R(\tilde x(k), \tilde u(k))$ a matrix of smooth functions with the property that 
\begin{eqnarray} \label{eq:convergence_d}
\lim_{{\tiny \begin{bmatrix} \tilde x(k) \\ \tilde u(k)\end{bmatrix}\to {0}}} R(\tilde x(k), \tilde u(k))= {0}
\end{eqnarray}
For $k=0$,  we have 
\[
\begin{bmatrix} \tilde x(0) \\ \tilde u(0) \end{bmatrix}=\varepsilon \begin{bmatrix} x(0) \\ u(0) \end{bmatrix},
\] 
hence, by  \eqref{eq:convergence_d},
$\lim_{\varepsilon\to {0}} R(\tilde x(0), \tilde u(0))= {0}$,
i.e. $R(\tilde x(0), \tilde u(0))$ converges to zero as fast as $\varepsilon$. 
The two just established facts and  \eqref{d.tilde} written for $k=0$ allow us to conclude that  $\tilde d(0)$ goes to zero faster than $\varepsilon$.  
We now proceed by induction. Let us assume that for some $k\ge 0$, the vectors $\begin{bmatrix} \tilde x(k) \\ \tilde u(k) \end{bmatrix}$ and $\tilde d(k)$ converge to zero 
at least as fast as $\varepsilon$
and faster than $\varepsilon$, respectively, 
and that $\lim_{\varepsilon\to {0}} R(\tilde x(k), \tilde u(k))= {0}$. Since
\[
\tilde x(k+1) = \begin{bmatrix} A  & B \end{bmatrix} \begin{bmatrix} \tilde x(k) \\  \tilde u(k)
\end{bmatrix} + \tilde d(k)
\]
then $\tilde x(k+1)$ converges to zero at least as fast as $\varepsilon$ and the same property holds for  $\begin{bmatrix} \tilde x(k+1) \\ \tilde u(k+1) \end{bmatrix}$ by design, since $\tilde u(k+1)=\varepsilon u(k)$. The identity \eqref{eq:convergence_d} written at time $k+1$ implies that $\lim_{\varepsilon\to {0}} R(\tilde x(k+1), \tilde u(k+1))= {0}$. By \eqref{d.tilde} written for $k+1$, we conclude that $\tilde d(k+1)$ converges to zero faster than $\varepsilon$ as claimed. \qedp

\medskip


Similar arguments can be used to show Assumption \ref{ass:dist2}. To this end,
recall from (\ref{eq:main1}) that $\tilde X_{1, T} = \varepsilon X^l_{1, T} + \tilde \Psi$,
where $X^l_{1, T}$ has full row rank. Thus, by applying the Young's inequality
we obtain 
\begin{eqnarray} 
\tilde X_{1, T} \tilde X_{1, T}^\top &=&  
(\varepsilon X^l_{1, T} + \tilde \Psi )  (\varepsilon X^l_{1, T} + \tilde \Psi )^\top \nonumber \\
& \succeq & \frac{\varepsilon^2}{2} X^l_{1, T} (X^l_{1, T})^\top - \tilde \Psi \tilde \Psi ^\top
\end{eqnarray} 
Hence, a sufficient condition for Assumption \ref{ass:dist2} to hold is that 
\[
\frac{\gamma \varepsilon^2}{2} X^l_{1, T} (X^l_{1, T})^\top
\succeq \gamma \tilde \Psi \tilde \Psi ^\top + \tilde D_{0, T} \tilde D_{0, T}^\top
\]
The result follows by noting that $X^l_{1, T}$ does not depend on $\varepsilon$ and 
is full row rank,
while the entries of $\tilde \Psi$ and $\tilde D_{0, T}$ are functions of $\tilde d$
which goes to zero faster than $\varepsilon$ (meaning that $\tilde \Psi \tilde \Psi ^\top$
and $\tilde D_{0, T} \tilde D_{0, T}^\top$ converge to zero faster than $\varepsilon^2$).

We summarize the results in the following theorem. 
Note that, by an abuse of notation, in the statement below as well as in the subsequent Theorem \ref{prop:nonlinear_stability_ass2}, when we refer to Assumptions \ref{ass:dist} and \ref{ass:dist2} and to problem (\ref{lmi.state.feedback.nonlinear}), we let the matrices $U_{0,1,T},X_{0,T},X_{1,T},D_{0,T}$ therein 
to be replaced by the matrices $\tilde{U}_{0,1,T},\tilde{X}_{0,T},\tilde{X}_{1,T},\tilde{D}_{0,T}$. 
\smallskip

\begin{theorem} \label{prop:nonlinear_stability_ass} 
Consider a nonlinear system as in \eqref{nonl}, with \textcolor{black}{state-space dimension $n$} and  
with equilibrium $(\overline x, \overline u)=(0,0)$, 
and let the corresponding linearized system be controllable. 
Consider any experiment $(x_0,u_{[0,T-1]})$ with persistently exciting
input $u_{[0,T-1]}$ of order $n+1$. 
Then, for any $\gamma>0$, there exists $\overline \varepsilon$
such that, for all $\varepsilon \in (0,\overline \varepsilon)$, the experiment $(\varepsilon x_0,\varepsilon u_{[0,T-1]})$
satisfies Assumptions \ref{ass:dist} and \ref{ass:dist2}. \qedp
\end{theorem}
\smallskip

\begin{figure} 
	\centering
	\psfrag{u(0)}{$u(0)$}
	\psfrag{u(1)}{$u(2)$}
		\includegraphics[width=0.5\textwidth]{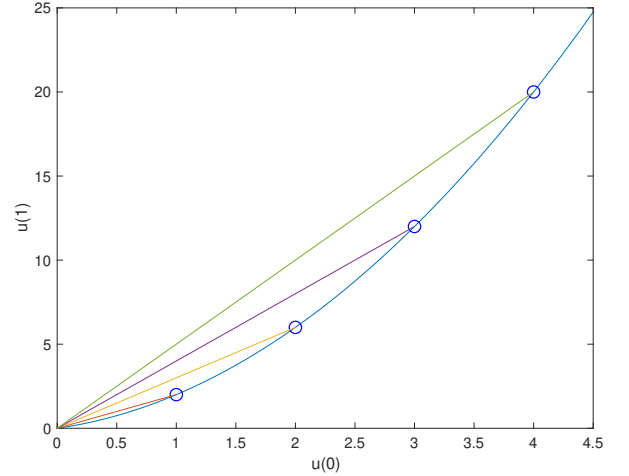}
	\caption{Pictorial representation of Theorem \ref{prop:nonlinear_stability_ass} 
	for the example of Section 3.1. The blue curve represents the set of points
        satisfying $[ u(0) \,\,\, u(1) ]  = [ \theta \,\,\, \theta + \theta^2 ]$ for which 
        Assumption \ref{ass:dist} fails to hold (for ease of illustration, we only report 
        the first two entries of the input).
        For each experiment (circles in the blue curve),
        the scaled versions $\varepsilon [ u(0) \,\,\, u(1) ]$ are represented by the coloured straight lines.}
        \label{fig:fig2}
\end{figure}

Theorem \ref{prop:nonlinear_stability_ass} gives a {\em principled} method 
to satisfy Assumptions \ref{ass:dist} and \ref{ass:dist2}.
In fact, if one can perform multiple experiments on the system at the equilibrium ($x_0=0$), one can 
apply scaled versions of the input sequence $u_{[0,T-1]}$ until Assumptions \ref{ass:dist} and \ref{ass:dist2}
are satisfied. 
In connection with the example of Section 3.1, an
intuitive explanation of Theorem \ref{prop:nonlinear_stability_ass} is given in Figure \ref{fig:fig2}.
\smallskip

We can further strengthen the result. 
\smallskip

\begin{theorem} \label{prop:nonlinear_stability_ass2} 
Consider a nonlinear system as in \eqref{nonl}, with \textcolor{black}{state-space dimension $n$} and  
with equilibrium $(\overline x, \overline u)=(0,0)$,
and let the corresponding linearized system be controllable. 
Consider any experiment $(\varepsilon x_0,\varepsilon u_{[0,T-1]})$ with persistently exciting
input sequence $u_{[0,T-1]}$ of order $n+1$. Then there exists $\overline \varepsilon$
such that, for all $\varepsilon \in (0,\overline \varepsilon)$:
\begin{enumerate}
\item[(i)]  Assumptions \ref{ass:dist} and \ref{ass:dist2} are satisfied;
\item[(ii)] problem (\ref{lmi.state.feedback.nonlinear})
is feasible;
\item[(ii)] the solution returns a stabilizing controller. 
\end{enumerate}
\qedp
\end{theorem}

\emph{Proof.} Point (i) has been previously shown. 

(ii). Take any experiment $(x_0,u_{[0,T-1]})$ with persistently exciting
input $u_{[0,T-1]}$.
If this experiment is applied to the linearized system, namely 
$(x^l_0,u^l_{[0,T-1]}) = (x_0,u_{[0,T-1]})$, we obtain from \cite[Theorem 3] {depersis-tesi2020tac} 
that
\begin{eqnarray}
\label{}
\begin{bmatrix}
X^l_{0,T} \,\overline Q &  X^l_{1,T} \overline Q \\
\overline Q^\top (X^l_{1,T})^\top & X^l_{0,T} \, \overline Q
\end{bmatrix}
\succ 0
\end{eqnarray}
for some matrix $\overline Q$. 
(We resort to Theorem 3 of \cite{depersis-tesi2020tac}  since it involves strict inequalities).

Recall now that for any scaled experiment $(\varepsilon x_0, \varepsilon u_{[0,T-1]})$ we have
\begin{eqnarray} \label{}
\tilde X_{0, T} = \varepsilon X^l_{0, T} + \tilde \Xi, \,\,
\tilde X_{1, T} = \varepsilon X^l_{1, T} + \tilde \Psi
\end{eqnarray} 
where $\tilde \Xi$ and $\tilde \Psi$ converge to zero faster than $\varepsilon$.
Thus, since $\overline Q$ is fixed, by taking $\varepsilon$ sufficiently small we obtain
\[
\begin{array}{rl}
\label{}
\begin{bmatrix}
\tilde{X}_{0,T} \,\overline Q &  \tilde{X}_{1,T} \overline Q \\
\overline Q^\top \tilde{X}_{1,T}^\top & \tilde{X}_{0,T} \, \overline Q
\end{bmatrix}
=&
\varepsilon
\begin{bmatrix}
{X}_{0,T}^l \,\overline Q &  {X}_{1,T}^l \overline Q \\
\overline Q^\top ({X}_{1,T}^l)^\top & {X}_{0,T}^l \, \overline Q
\end{bmatrix}\\
&
+
\begin{bmatrix}
\tilde{\Xi} \,\overline Q &  \tilde{\Psi} \overline Q \\
\overline Q^\top \tilde{\Psi}^\top & \tilde{\Xi} \, \overline Q
\end{bmatrix}
\succ 0
\end{array}
\]
This also implies that 
\begin{eqnarray}
\label{}
\begin{bmatrix}
\tilde{X}_{0,T} \,\overline Q - \alpha \tilde{X}_{1,T}  \tilde{X}_{1,T}^\top &  \tilde{X}_{1,T} \overline Q \\
\overline Q^\top  \tilde{X}_{1,T}^\top & \tilde{X}_{0,T} \, \overline Q
\end{bmatrix}
\succ 0
\end{eqnarray}
for sufficiently small $\alpha$. This shows the feasibility of the first of (\ref{lmi.state.feedback.nonlinear}).

Regarding the second of (\ref{lmi.state.feedback.nonlinear}),
first notice that the above LMI remains satisfied if we scale it by $\delta > 0$, that is
\begin{eqnarray}
\label{7a.scaled}
\begin{bmatrix}
\tilde{X}_{0,T} \, \hat Q - \delta \alpha \tilde{X}_{1,T}  \tilde{X}_{1,T}^\top &  \tilde{X}_{1,T} \hat Q \\
\hat Q^\top \tilde{X}_{1,T}^\top & \tilde{X}_{0,T} \, \hat Q
\end{bmatrix}
\succ 0, \quad \hat Q := \delta \overline Q
\end{eqnarray}

for all $\delta > 0$.
Consider now the matrix 

\[
I_T - \hat Q (\tilde{X}_{0,T} \,\hat Q)^{-1} \hat Q^\top  
\]

which is well defined since $\tilde{X}_{0,T} \,\hat Q \succ 0$. 
Exploiting the relation $\hat Q := \delta \overline Q$,
we obtain 
\[
I_T - \hat Q (\tilde{X}_{0,T} \,\hat Q)^{-1} \hat Q^\top   =
I_T - \delta \overline Q (\tilde{X}_{0,T} \,\overline Q)^{-1}  \overline Q^\top  \succ 0
\]
for sufficiently small $\delta$. By the Schur complement,
this shows the feasibility of the second of (\ref{lmi.state.feedback.nonlinear}),
which together with \eqref{7a.scaled}  concludes the proof of (ii).

(iii). Here we exploit the fact that  
in the design formulation we search for the solution maximizing $\alpha$. 
Let $(\overline Q, \overline \alpha)$ be the solution to (\ref{lmi.state.feedback.nonlinear})
associated to the experiment $(x_0,u_{[0,T-1]})$ carried out on the linearized system,
with corresponding matrices $X^l_{0,T}$ and $X^l_{1,T}$.
Consider the scaled experiment $(\varepsilon x_0,\varepsilon u_{[0,T-1]})$ still 
on the linearized system, resulting in state matrices $\varepsilon X^l_{0,T}$ and $\varepsilon X^l_{1,T}$.
Denote its solution by $(Q, \alpha)$.
It follows that $\alpha = \overline \alpha$, that is the optimal value of $\alpha$ does not change. 
To see this fact, we first show 
that $(Q, \alpha) = (\varepsilon \overline Q, \overline \alpha)$ is feasible. 
For the first of (\ref{lmi.state.feedback.nonlinear}), 
the scaled experiment $(\varepsilon x_0,\varepsilon u_{[0,T-1]})$ gives
 \begin{eqnarray}
\label{}
\begin{bmatrix}
\varepsilon X^l_{0,T} \, Q - \varepsilon^2 \alpha X_{1,T}^l  (X_{1,T}^l)^\top &  \varepsilon X_{1,T}^l  Q \\
\varepsilon Q^\top (X_{1,T}^l)^\top & \varepsilon X_{0,T}^l \,  Q
\end{bmatrix}
&=& \nonumber \\
 \begin{bmatrix}
 \varepsilon^2 X^l_{0,T} \, \overline Q - \varepsilon^2 \overline \alpha X_{1,T}^l  (X_{1,T}^l)^\top &   
 \varepsilon^2 X_{1,T}^l  \overline Q \\
\varepsilon^2 \overline Q^\top (X_{1,T}^l)^\top & \varepsilon^2 X_{0,T}^l \,  \overline Q
\end{bmatrix} \succeq 0
\end{eqnarray}
since $(\overline Q, \overline \alpha)$
is a feasible solution to (\ref{lmi.state.feedback.nonlinear})
associated with the experiment $(x_0,u_{[0,T-1]})$.

Concerning the second of (\ref{lmi.state.feedback.nonlinear}), 
the scaled experiment $(\varepsilon x_0,\varepsilon u_{[0,T-1]})$ gives
\begin{eqnarray} \label{second.of.(7)}
\begin{bmatrix}
I_T &  Q \\
 Q^\top & \varepsilon X_{0,T}^l \,Q
\end{bmatrix}
=
\begin{bmatrix}
I_T &  \varepsilon \overline Q \\
\varepsilon \overline Q^\top & \varepsilon^2 X_{0,T}^l \,\overline Q
\end{bmatrix}  \succeq 0
\end{eqnarray}
since $(\overline Q, \overline \alpha)$
is a feasible solution to (\ref{lmi.state.feedback.nonlinear})
associated with $(x_0,u_{[0,T-1]})$.
In fact, by the Schur complement, \eqref{second.of.(7)} holds if and only if 
$I_T-  \overline Q (X_{0,T}^l \,\overline Q)^{-1} \overline Q^\top\succeq 0$.

We finally show that for the scaled experiment 
we cannot have a solution with $\alpha > \overline \alpha$. Consider the solution $(Q, \alpha)$
associated with the scaled experiment $(\varepsilon x_0,\varepsilon u_{[0,T-1]})$
and suppose by contradiction that $\alpha > \overline \alpha$. 
We show that in this case $(\hat Q, \hat \alpha) := (Q/\varepsilon, \alpha)$
is a solution associated with the experiment $(x_0,u_{[0,T-1]})$, contradicting the 
fact that $\overline \alpha$ is the optimal value. For the 
first of (\ref{lmi.state.feedback.nonlinear}) the solution $(Q, \alpha)$ 
associated with the scaled experiment $(\varepsilon x_0,\varepsilon u_{[0,T-1]})$ gives
 \begin{eqnarray}
\label{}
\begin{bmatrix}
 \varepsilon X^l_{0,T} \, Q -  \alpha \varepsilon^2 X_{1,T}^l  (X_{1,T}^l)^\top & \varepsilon  X_{1,T}^l   Q \\
\varepsilon Q^\top (X_{1,T}^l)^\top &  \varepsilon X_{0,T}^l \,  Q
\end{bmatrix}
\succeq 0
\end{eqnarray}
From the relation $(\hat Q, \hat \alpha) = (Q/\varepsilon, \alpha)$ 
we would thus get
 \begin{eqnarray}
\label{}
 \begin{bmatrix}
 \varepsilon^2 X^l_{0,T} \, \hat Q - \varepsilon^2 \hat \alpha X_{1,T}^l  (X_{1,T}^l)^\top &   
 \varepsilon^2 X_{1,T}^l  \hat Q \\
\varepsilon^2 \hat Q^\top (X_{1,T}^l)^\top & \varepsilon^2 X_{0,T}^l \,  \hat Q
\end{bmatrix} \succeq 0
\end{eqnarray}
showing that $(\hat Q, \hat \alpha)$ with $ \hat \alpha > \alpha$ is a solution to
the first of (\ref{lmi.state.feedback.nonlinear}) for the experiment $(x_0,u_{[0,T-1]})$.
Similarly, for the second of (\ref{lmi.state.feedback.nonlinear}) 
the solution $(Q, \alpha)$ 
associated with the scaled experiment $(\varepsilon x_0,\varepsilon u_{[0,T-1]})$ gives
\begin{eqnarray} \label{}
\begin{bmatrix}
I_T &  Q \\
 Q^\top & \varepsilon X_{0,T}^l \,Q
\end{bmatrix} \succeq 0
\end{eqnarray}
From the relation $(\hat Q, \hat \alpha) = (Q/\varepsilon, \alpha)$ 
we would thus get
\begin{eqnarray} \label{}
\begin{bmatrix}
I_T &  Q \\
 Q^\top & \varepsilon X_{0,T}^l \,Q
\end{bmatrix}
=
\begin{bmatrix}
I_T &  \varepsilon \hat Q \\
\varepsilon \hat Q^\top & \varepsilon^2 X_{0,T}^l \,\hat Q
\end{bmatrix} \succeq 0
\end{eqnarray}

To summarize, for the linearized system all the experiments $(\varepsilon x_0,\varepsilon u_{[0,T-1]})$
result in the same optimal value for $\alpha$. The proof is concluded noting that for the nonlinear system,
the state matrices $X_{0,T}$ and $X_{1,T}$ 
associated with the experiment $(\varepsilon x_0,\varepsilon u_{[0,T-1]})$
converge to $\varepsilon X^l_{0,T}$ and $\varepsilon X^l_{1,T}$, respectively, as $\varepsilon$ 
converges to zero, and this implies that
the optimal value of $\alpha$ for the nonlinear system converges to $\overline \alpha$. 
In turn,
this implies that for $\varepsilon$ sufficiently small the solution returns a stabilizing controller
since: (a) $\overline \alpha$ is fixed; and, as stated in Theorem \ref{prop:nonlinear_stability_ass}, 
(b) for any given $\gamma$, 
Assumption \ref{ass:dist2} is satisfied for $\varepsilon$ sufficiently small. 
\qedp

\subsection{Practical considerations}

The foregoing analysis also gives a simple method to assess whether the solution 
returns a stabilizing controller. Specifically, if we have some prior knowledge on 
$D_{0, T}$, that is on the type of nonlinearity, then we can directly assess the fulfilment of the
condition $\gamma < \alpha^2/(4 + 2\alpha)$ by computing the smallest value of $\gamma$
that satisfies Assumption \ref{ass:dist2}. If instead we have no prior knowledge on 
$D_{0, T}$, one can perform multiple experiments on the system, each one being
an $\varepsilon$-scaled version of the experiment $(\varepsilon x_0,\varepsilon u_{[0,T-1]})$. 
In this case, it follows from the foregoing analysis that $\alpha$ converges (faster than linearly) 
to the optimal \emph{fixed} value $\overline \alpha$ associated with solution for the linearized system.

\begin{table*}[t!]
\centering
\begin{tabular}[b]{|c|c|c|c|c|}
\hline
& $\| K - \overline K \|$ & $|\alpha - \overline \alpha| $ & 
Stability achieved & $\gamma < \alpha^2/(4 + 2\alpha)$ \\
&  &  & & fulfilled \\
\hline
\hline
linear system & 0  & 0  & YES & YES \\
($\varepsilon$-invariant) &  &  & &  \\
\hline
nonlinear system & 0.0082  & 2e-7  & YES & YES \\
($\varepsilon =1$) &  &  & &  \\
\hline
nonlinear system & 0.0020  & 5e-8  & YES & YES \\
($\varepsilon =0.5$) &  &  & &  \\
\hline
nonlinear system & 0.0001  & 2e-9  & YES & YES \\
($\varepsilon =0.1$) &  &  & &  \\
\hline
nonlinear system & 1e-5  & 3e-10  & YES & YES \\
($\varepsilon =0.01$) &  &  & &  \\
\hline
\hline
\end{tabular}
\caption{Some tests for the inverted pendulum system.}
\label{tab:relationship}
\end{table*}

We illustrate these considerations through a numerical example.
Consider the Euler discretization of an inverted pendulum
\[
\begin{array}{rl}
x_1(k+1) =& x_1(k)+ \Delta x_2(k) \\[0.5cm]
x_2(k+1) =& \displaystyle  \frac{\Delta  g}{\ell} \sin x_1(k) + \left( 1 - \frac{\Delta  \mu}{m \ell^2} \right) x_2(k) 
+\frac{\Delta }{m \ell^2}u(k)\\
\end{array}
\]
where we simplified the sampled times $k \Delta$ in $k$, with $\Delta$ the sampling time. 
In the model $m$ is the mass to be balanced, $\ell$ is the distance from the base to the center of mass of the
balanced body, $\gamma$ is the coefficient of rotational friction, $g$ is the acceleration due to gravity. The states $x_1, x_2$ are the angular position and velocity, respectively, $u$ is the applied torque.
The system has an unstable equilibrium in 
$(\overline x, \overline u)=(0,0)$ corresponding to the pendulum upright position.

The first experiment is generated starting at the equilibrium ($x_0=0$),
with random input sequence within the interval $[-5,5]$, which causes 
a displacement of about $28$ degrees from the equilibrium position.
All the other experiments 
are generated starting at the equilibrium and by scaling the input sequence of a factor $\varepsilon$.
The solution for the linearized system results in $\overline \alpha = 0.01136$
and $\overline K = [-56.9163 \,\,\, -12.6186]$. (Notice that we allow for such large input sequences since
we start from $x_0=0$.)
 
As the examples shows, $\alpha$ converges to $\overline \alpha$ very quickly.
In this respect, an interesting point observed in simulations is that also $K$ converges to 
$\overline K$. In fact, simulations indicate that the solution $Q$ for the nonlinear 
system with experiment $(\varepsilon x_0,\varepsilon u_{[0,T-1]})$ converges to
$\varepsilon \overline Q$, where $\overline Q$ is the solution for the linearized system 
with experiment $( x_0, u_{[0,T-1]})$. This property immediately gives that 
$K = U_{0,1,T} Q (X_{0,T} Q)^{-1}$ converges to
$\varepsilon U_{0,1,T} \varepsilon \overline Q (\varepsilon X^l_{0,T} \varepsilon \overline Q)^{-1}
= \overline K$.


\section{Conclusions}\label{sec:conclusion}

We have studied the conditions under which a controller that stabilizes unknown nonlinear systems in the first approximation directly from experimental data can be designed. We have shown via an example that these conditions may not hold even when the data have been obtained applying persistently exciting inputs. This is in sharp contrast with the results derived for linear systems. Nevertheless, for those experiments for which the conditions for the existence of a stabilizer of the nonlinear system do not hold, we have shown that a suitable scaling of the initial conditions and the applied input leads to the fulfilment of these conditions, and hence they allow the design of a data-based stabilizer of the unknown nonlinear system. We regard these results as a principled method to design stabilizers in the first approximation and will also have an impact on the nonlocal design of stabilizers for nonlinear systems, which we are currently investigating. Important practical aspects have been neglected. The scaling of the experiments may make the obtained data more sensitive to noise. Although the approach of \cite{depersis-tesi2020tac} allows for a robust analysis of the effect of noise on the controller design, this effect has not been studied in the current paper and deserves attention. 
%

\bibliography{biblio-data}             

\vspace{0.5cm}

%
%

\end{document}